\documentstyle[preprint,prd,aps,floats]{revtex}
\input{psfig.tex}
\begin{document} 
\draft
\preprint{
\begin{tabular}{rr}
CfPA/96-th-15\\
\end{tabular}
}

\title{How Anisotropic is our Universe?}
\author{Emory. F. Bunn, Pedro Ferreira and Joseph Silk }
\address{Center for Particle Astrophysics,  and  Departments of Astronomy and Physics,\\
University of California, Berkeley  CA 94720-7304}
\maketitle
\begin{abstract}
Large-scale cosmic microwave background anisotropies
in homogeneous, globally  anisotropic cosmologies are investigated.
We perform a statistical analysis in which  the four-year
data from the
Cosmic Background Explorer
satellite is searched for the specific anisotropy patterns predicted by these models
and thereby
set definitive upper limits on the amount of shear, 
$(\sigma/H)_0$ and vorticity, $(\omega/H)_0$, which are
orders of magnitude stronger than previous constraints. We comment on
how these results might impact our understanding of primordial global
anisotropy.
\end{abstract}

\date{\today}


\renewcommand{\thefootnote}{\arabic{footnote}}
\setcounter{footnote}{0}
Fluctuations in the cosmic microwave background radiation
(CMBR) provide us with a clean and unique probe of the structure
of our universe on both small and large scales \cite{Review}.
Current experiments  are allowing us to learn 
about processes in the early universe with unparalleled precision,
and the prospect of future terrestrial and satellite missions
give us reason to hope for a
clear and accurate picture of
the universe in the near future.
The main working hypothesis in our attempts to understand 
the universe is that we live in what was originally a 
homogeneous and isotropic spacetime, the 
Friedmann-Robertson-Walker (FRW) cosmology. The current belief is that
some physical mechanism generated perturbations (either
primordial, with inflation, or actively, with defects) that
evolved through gravitational collapse to form 
the structures we see now. The  smoothness of the
CMBR seems to be consistent with such a picture.

Inflation provides the primary motivation for believing that,
at least in an initially homogeneous space-time, any primordial anisotropy has been removed.
However, inflation is by no means a generic initial condition for the universe. 
In the absence of inflation, one is likely to have commenced with global metric perturbations that while eventually seeding structure formation may result in a universe that is only asymptotically Friedmann-Robertson-Walker.
Bianchi models provide a generic description of homogeneous anisotropic cosmologies. In the spirit of studying an alternative to inflation,
we study below the experimental
constraints if our universe  is only asymptotically FRW.
 There are distinct features depending on the overall geometry and
homogeneity class of the model \cite{Bianchi}, 
and in a pioneering paper,  Collins
and Hawking  used analytical arguments to find upper bounds 
on the amount of shear ($\sigma_0$) and vorticity ($\omega_0$)
in the universe today, from the absence of any detected
 CMBR anisotropy. A detailed numerical analysis of such models \cite{BJS}  
used experimental limits on the dipole and quadrupole to refine
limits on universal rotation. More recently, Barrow has
argued \cite{Barrow} that there is no ``isotropy problem'' in  such cosmological models
which are maximally anisotropic (i.e. in which shear and vorticity
have decayed only logarithmically since the Planck time):
if one assumes an equipartition of energy 
between all modes of the gravitational interaction at the Planck
time then the present amplitude of CMBR fluctuations should be
compatible with current observational limits.

In this {\it Letter} we  improve on previous bounds
on the total shear in the universe. We use the most
recent data from the Differential Microwave Radiometers
aboard the Cosmic Background Explorer (COBE) \cite{bennett} to
constrain the allowed parameters of a Bianchi model of
type VII$_h$; this model is asymptotically close to an
open FRW universe and has the richest anisotropy structure
of the models we could consider. As pointed out by Barrow,
it is also an example of a homogeneous cosmology where 
the decay in $\sigma$ from the Planck time to the present
is minimal.

One can describe Bianchi cosmologies in terms of the metric
\begin{eqnarray}
g_{\mu\nu}=-n_\mu n_\nu+a^2[\exp(2\beta)]_{AB}e^A_\mu e^B_\nu,
\end{eqnarray}
where $n_\alpha$ is the normal to spatial hypersurfaces of
homogeneity, $a$ is the conformal scale factor, $\beta_{AB}$ is a 
$3 \time 3$ matrix only dependent on
cosmic time, $t$, and $e^A_\mu$ are invariant covector
fields on the surfaces of homogeneity, which obey the commutation
relations
\begin{eqnarray}
e^A_{\mu;\nu}-e^A_{\nu;\mu}=C^A_{BC}e^B_{\mu}e^C_{\nu}.
\end{eqnarray}
The structure constants
$C^A_{BC}$ can be used to classify the different models.
We shall focus on type VII$_h$ which has structure constants
$C^2_{31}=C^3_{21}=1,\  C^2_{21}=C^3_{31}=\sqrt{h}$.
It is convenient to define the parameter
$x=\sqrt{{h/{1-\Omega_0}}}$, which determines the scale on which
the principal axes of shear and rotation change orientation.
By taking combinations of limits of $\Omega$ and $x$ one can 
obtain Bianchi I, V and VII$_0$ cosmologies. 

We are interested in large-scale anisotropies so it suffices
to evaluate the peculiar redshift a photon will feel from
the epoch of last scattering (${\it ls}$) until now (0)
\begin{eqnarray}
{\Delta T_A}({\bf {\hat r}})=({\hat r}^iu_i)_{0}
-({\hat r}^iu_i)_{\it ls}-\int^{0}_{\it ls}{\hat r}^j{\hat r}^k
\sigma_{jk}d\tau 
\label{eq:dt1}
\end{eqnarray}
where ${\bf {\hat r}}=
(\cos\theta\sin\phi,\sin\theta\sin\phi,\cos\phi)$ 
is the direction vector of the incoming null
geodesic, ${\bf u}$ is the spatial part of the fluid four-velocity
vector and to first order, the shear is 
$\sigma_{ij}={\partial_{\tau}} \beta_{ij}$. To evaluate expression 
(\ref{eq:dt1}),
one must first of all determine a parameterization of geodesics
on this spacetime. This is given by
\begin{eqnarray}
\tan({{\phi(\tau)} \over 2}) &=&\tan({\phi_0\over 2})
\exp[-(\tau-\tau_0)\sqrt{h}] \nonumber \\
\theta(\tau)&=&\theta_0+(\tau-\tau_0) \nonumber \\-
{1 \over \sqrt{h}}
\ln\{\sin^2({\phi_0\over 2})&+&\cos^2({\phi_0\over 2})
\exp[2(\tau-\tau_0)\sqrt{h}]\}
\end{eqnarray}
Solving Einstein's equations (and assuming that matter is
a pressureless fluid)
one can determine ${\bf u}$ and $\sigma_{ij}$.
A general expression for (\ref{eq:dt1}) was determined in 
\cite{BJS}:
\begin{eqnarray}
\begin{array}{l}
{\Delta T}_A({\bf {\hat r}})=({\sigma \over H})_0{2\sqrt{1-\Omega_o} 
\over {\Omega_0}} \nonumber \\
\times \{ [\sin\phi_0\cos \theta_0-\sin \phi_{ls}
\cos \theta_{ls}(1+z_{ls})] \nonumber \\
- \int_{\tau_{ls}}^{\tau_0}{{3h(1-\Omega_0)} \over \Omega_0}
\sin 2\phi [\cos(\theta)+\sin(\theta)]{{d\tau} \over {\sinh^4 (
\sqrt{h}\tau/2)}}\} \nonumber 
\end{array}\\ \;\label{eq:defTA}
\end{eqnarray}
As shown in \cite{Bianchi},\cite{BJS}, 
the ``patterns'' in such a model are easy to describe:
for $\Omega_0 < 1$ and a finite $x$ one will obtain
a spiral with approximately $N=2/\pi x$ complete twists,
focused towards the axis of rotation with an angular size of 
order $1 /\Omega_0$.
Taking $x \rightarrow \infty$ will leave only a hotspot.

We will approach the problem of constraining the parameters of these
Bianchi models in the following way.  For fixed values of the parameters
$x$ and $\Omega_0$, we will attempt to place upper limits on the amplitude
of the shear $(\sigma/H)_0$ (or equivalently vorticity $(\omega/H)_0$
\cite{vort}).
  The statistics problem we face
differs substantially from the situation encountered in placing
constraints on more standard
models.  In standard cosmological models, the predicted CMB anisotropy
is a realization of an isotropic Gaussian random field, and its
statistical properties are therefore entirely characterized by the
power spectrum $C_l$.
In contrast, in the Bianchi models at least
part of the CMB fluctuation comes from the large-scale anisotropy of
space; this contribution to the anisotropy takes the form of a
particular pattern on the sky, and is not described by the statistics
of a Gaussian random field.  We therefore require different
statistical techniques from those used in previous analyses.

Let $\Delta T({\hat{\bf r}})$ be the temperature fluctuation in the
direction of 
the unit vector ${\hat{\bf r}}$.  We assume that $\Delta T$ is the sum of two
contributions: 
\begin{eqnarray}
\label{DeltaTDef}
\Delta T({\hat{\bf r}}) = \Delta T_A({\hat{\bf r}}) + \Delta T_I({\hat{\bf r}}).
\end{eqnarray}
Here $\Delta T_A$ is defined in (\ref{eq:defTA}) and $\Delta T_I$ 
represents the ``isotropic''
residual fluctuation caused by variations in the density and
gravitational potential.
We call $\Delta T_I$ ``isotropic'' because we
assume that it is described by the statistics of an isotropic
Gaussian random field.  That is, if we expand it in spherical harmonics,
\begin{eqnarray}
\Delta T_I({\hat{\bf r}}) = \sum_{l,m}a_{lm}Y_{lm}({\hat{\bf r}}),
\end{eqnarray}
then the coefficients $a_{lm}$ are independent Gaussian random variables
with zero mean and variances given by an angular power spectrum $C_l$:
$\langle a_{lm}a^*_{l'm'} \rangle = C_l\delta_{ll'}\delta_{mm'}$.

One can naively assume that this set of Gaussian perturbations
was generated through the amplification of quantum fluctuations,
as in the case of inflation (however see \cite{isotrop}) and 
it is quite conceivable that this initial set of perturbations
is strongly biased (or anti-biased) due to the large
anisotropy at early times. To change our results considerably,
this would have to compensate (anticompensate) late
time evolution of the overall anisotropy  on many scales, i.e.
there would have to be a strong correlation between the primordial
{\it quantum} generation of perturbations and subsequent {\it classical}
evolution of the different temperature variables such that they 
would interfere (destructively or constructively) for many modes of
the temperature autocorrelation function,
 something
that we believe to be unlikely.
Another possibility is that subsequent evolution of perturbations
will be locked in to the specific orientation of the large scale
anisotropy, but if we assume that the dominant source of
perturbations are scalar and therefore are only sensitive to
the overall volume change of the spacetime, than we can discard
this hypothesis. This reasoning does not hold if the dominant
source of perturbations comes from tensor modes, but we shall
not consider this possibility here.

The anisotropic component $\Delta T_A$ does not
obey Gaussian statistics.  Rather, for fixed values of $x$, $\Omega_0$,
and $\sigma$, we can compute the exact pattern $A({\hat{\bf r}})$ of the
CMB anisotropy.  Unfortunately, we do not know the orientation of
this pattern on the sky.  We can say that
\begin{eqnarray}
\Delta T_A = A({\bf R}{\hat{\bf r}}),
\end{eqnarray}
where $A$ is the known pattern of anisotropy and ${\bf R} \in SO(3)$ is a
$3\times 3$ rotation matrix.  The matrix $\bf R$ can be specified
by three Euler angles $(\theta,\psi,\varphi)$, but we have no knowledge
{\it a priori} of the values of these angles.


We will place constraints on the shear following standard frequentist
statistical practice.  (This is in contrast to the Bayesian philosophy
adopted in much cosmological data analysis.)  For any particular
model, we define some goodness-of-fit statistic $\eta$ that depends
on the data 
(To give a simple, familiar example, when one is trying to estimate the
mean of a set of data, it is customary to choose $\eta$ to be the 
chi-squared of the data).
Having chosen our goodness-of-fit statistic, we compute its value
$\eta_*$ using the actual data.  We then compute the probability
$P(\eta < \eta_*)$ that a random data set would have produced a value
as good a value as the actual data or better.  If this probability is
large, then we say that the model is inconsistent with the data.  It
is customary to choose a significance level $P_0$, say $0.95$, and say
that a particular model is ruled out at that significance level if
$P(\eta < \eta_*) > P_0$.

Our choice of $\eta$ is as follows. Each pixel $d_i$ of our data
set contains contributions from both intrinsic CMB anisotropy and
noise:
\begin{eqnarray}
d_i = (\Delta T\star B)({\hat{\bf r}}_i) + N_i.
\label{COBEDataEq}
\end{eqnarray}
here $B$ represents the DMR beam pattern \cite{wright}, 
${\hat{\bf r}}_i$ is the direction
on the sky of the $i$th data point, and the star denotes a convolution.
According to our model, $\Delta T$ includes the two contributions
shown in equation (\ref{DeltaTDef}).  $N_i$ is the noise in pixel $i$.
For the COBE data, it is an excellent approximation to take $N_i$ to
be Gaussian with zero mean.  The correlations between the
noise in different pixels are negligible \cite{lineweaver}, so
$\langle N_i N_j\rangle = \sigma_i^2\delta_{ij}$.

Before we give the actual definition of the statistic $\eta$, let
us consider a simpler case.
Suppose that we knew the geometrical parameters $x$, $\Omega_0$, and $\sigma$,
{\it and} the rotation matrix $\bf R$ that defines the orientation of the
pattern $A$ on the sky.  Then the anisotropic part $\Delta T_A$ of the
CMB fluctuation would be completely specified, but the isotropic portion
$\Delta T_I$ and the noise $N_i$ would be completely unknown.  In this
situation, we could define a natural goodness-of-fit statistic in the
following way.  Let $\Delta_0^2$ be the noise-weighted 
mean-square value of the data:
\begin{eqnarray}
\Delta_0^2 = \sum_i{d_i^2\over\sigma_i^2}.
\end{eqnarray}
Now let $\Delta_1^2$ be the mean-square value of the residuals
after we have subtracted off the known anisotropic portion:
\begin{eqnarray}
\Delta_1^2 = \sum_i {\left(d_i - (\Delta T_A\star B)({\hat{\bf r}}_i)\right)^2
\over \sigma_i^2}.
\end{eqnarray}
If our model is correct, then we expect $\Delta_1^2$ to be smaller
than $\Delta_0^2$: if we have correctly removed a portion of the
signal, then the residuals should be smaller, on average.  On the
other hand, if our model is incorrect, then attempting to remove the
anisotropic portion should {\it increase} the residuals.  The
difference between $\Delta_0^2$ and $\Delta_1^2$ is therefore a natural
choice of goodness-of-fit statistic.  In practice, 
it is more convenient
to divide by $\Delta_0^2$, in order to make the results more weakly
dependent on the amplitude of the isotropic cosmological signal $\Delta T_I$.
We therefore define our statistic to be
\begin{eqnarray}
\label{etaSimple}
\eta_1={\Delta_0^2-\Delta_1^2\over\Delta_0^2}.
\end{eqnarray}

In fact, of course, we cannot use the statistic (\ref{etaSimple}),
because we do not know the parameters necessary to determine $\Delta
T_A$.  In particular, we do not know $\bf R$ or $\sigma$.  (We have
chosen to set up the problem as one of constraining $\sigma$ for fixed
$x$ and $\Omega_0$, so we can assume that we know the latter two
parameters.)  But we can define a new statistic $\eta$ whose
value is the minimum of $\eta_1$ over the unknown parameters:
\begin{eqnarray}
\label{etaEq}
\eta = \min_{\sigma,\theta,\psi,\varphi}\eta_1.
\end{eqnarray}

The statistical task we have set ourselves is simple in principle,
although it is somewhat cumbersome computationally.  For fixed
values of the parameters $x$ and $\Omega_0$, we must compute the value
$\eta_*$ of the statistic (\ref{etaEq}) for the real data.  We must
then perform Monte Carlo simulations to
determine the theoretical probability distribution of $\eta$ to
see how consistent the actual value is with each theoretical model.
We must perform these simulations for a variety of different values
of the shear $\sigma$ in order to see which values of $\sigma$ are
consistent with the data.

Each calculation of $\eta$ involves a minimization in a
four-dimensional parameter space.  Since we need to compute $\eta$
repeatedly in our simulations, it is important to perform this
calculation efficiently.  We chose to reduce the numbers of pixels in
the COBE data set by binning pixels together in groups of four ({\it
i.e.}, working in ``pixelization level 5'' rather than level 6).
Since the anisotropy pattern $\Delta T_A$ tends to have power on
larger scales than either the noise or the isotropic signal, this
binning does not reduce our sensitivity very much.

The task of finding the minimum in equation (\ref{etaEq}) is not
trivial, since the function $\eta_1(\sigma,\theta,\psi,\varphi)$ has
numerous local minima.  We use Powell's method for finding minima, but
we have to try multiple starting points in order to be confident that
we had found, if not the true minimum, at least a local minimum that
was almost as low as the true minimum.  We chose to adopt the
following procedure.  We choose $p$ random points in parameter space
and evaluate $\eta_1$ at each.  Starting from the point that gave the
lowest value of $\eta_1$, we use Powell's method to find a local
minimum.  We then repeat this entire procedure $q$ times, and we take
the lowest value found to be our statistic $\eta$.  After some
experimentation, we found that choosing $p=10$ and $q=4$ gave
reasonable results.  Of course, it is essential to use precisely the
same procedure for determining $\eta$ in both the real data and the
simulations.

We perform our analysis on the four-year DMR data set \cite{bennett}.  
We use the
ecliptic pixelization of the data.  Before performing any analysis, we
average together the two 53 GHz maps and the two 90 GHz maps to make a
single sky map.  The averaging is performed with weights inversely
proportional to the squares of the noise levels, in order to minimize
the noise in the average map.  In order to reduce Galactic
contamination, we excise all pixels that lie within the ``custom cut''
described by the COBE group \cite{kogut}; this reduces the number of
pixels in the map from 6144 to 3890.  We then remove a best-fit
monopole and dipole from the map.  As mentioned above, we degrade
the map from pixelization level 6 to level 5, and compute the statistic
$\eta$ for a grid of points in the $\Omega_0$-$x$ plane.

In order to convert these $\eta$ values into constraints on the
shear, we need to determine the probability distribution of $\eta$
via Monte Carlo simulations.  We performed simulations on a
sample of 10 models in the $\Omega_0$-$x$ plane, using
three or four values of $\sigma$ for each model.  For each choice
of the three parameters $(\Omega_0,x,\sigma)$, we created between
200 and 500 random DMR data sets according to equation (\ref{COBEDataEq}).
For simplicity,
we assumed that the isotropic component $\Delta T_I$ of the anisotropy
was given by a scale-invariant power spectrum $C_l^{-1}\propto l(l+1)$,
although our final results are not very sensitive to this assumption.
[Specifically, if
we steepen the power spectrum to an effective $n=1.5$ spectrum
(see, {\it e.g.}, \cite{Review} for a definition), the limits
in Figure 1 change by $\sim 20\%$.]
We processed each sky map in the same way as the real data to
determine a value of $\eta$.

We found that in every case the probability distribution of $\eta$ was
slightly skew-positive and had tails that were consistent with
exponential distributions.  We determined the first three moments of
each probability distribution, and found that each distribution was
very well approximated by a stretched, offset chi-squared
distribution, where the three parameters of the distribution (stretch,
offset, and number of degrees of freedom) were chosen to fit the three
moments \cite{tegmarkbunnhu}.  
For points in parameter space
where we have not performed simulations, we assume that the probability
distribution is also well approximated by a stretched chi-squared
distribution, and we determine the three parameters of the distribution
by smoothly interpolating between the points where we have 
performed simulations.

Having estimated the probability distribution of $\eta$ for the
various theoretical models in this way, we are able to set limits
on the shear.  For each point in our grid in the $\Omega_0$-$x$ plane,
we determine the range of values of $\sigma$ such that $P(\eta<\eta_*)<0.95$.
We find that $\sigma=0$ is always allowed at the 95\% confidence level;
{\it i.e.}, we do not detect shear at this level.  Figure 1
shows the upper limits we can set on $(\sigma/H)_0$ and $(\omega/H)_0$
as a function of $\Omega_0$
and $x$: for $\Omega_0=1$ universes we find that $(\sigma/H)_0<3\times
10^{-9}$ (or $(\omega/H)_0<10^{-6}$) for any $x>.05$, while for
$\Omega_0<1$ the upper bounds are even tighter. 

\begin{figure}[t]	
\centerline{\psfig{file=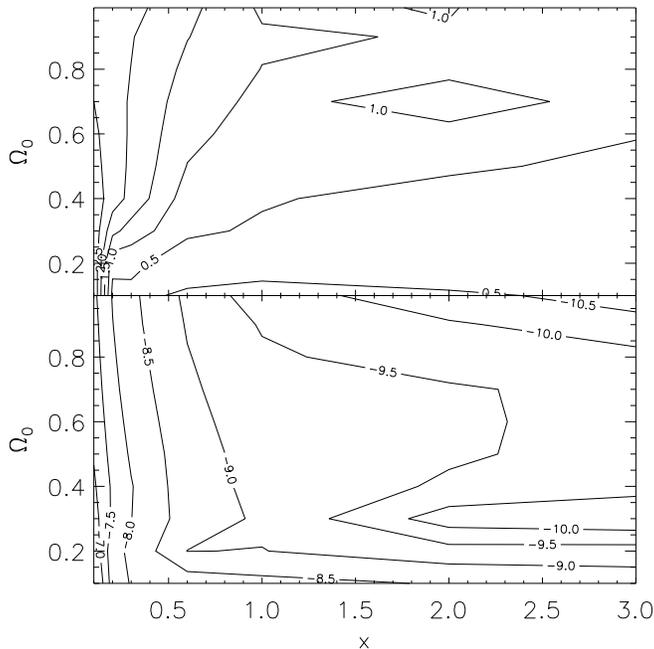,width=3.5in}}
\caption{Upper bounds on shear and vorticity: contours of equal
$(\sigma/H)_0\times 10^9$ and
$\log_{10}(\omega/H)_0$ are shown in the upper and lower panels
respectively for a class of Bianchi
VII$_h$ models.
}
\end{figure}

These values are to be
compared with the constraints from \cite{BJS} which are typically one
to two orders of magnitude higher and relied entirely on the quadrupole:
at that time  $Q\simeq7\times10^{-5}K$ compared to $Q\simeq1-2\times
10^{-5}K$ from the COBE DMR data \cite{bennett}. 
Moreover, in discarding information from higher moments,
 their comparison was not sensitive to the small-scale structure
present in anisotropic models that is associated  either with the spiral 
pattern (which  introduces power on smaller scales as x
increases) or with geometrical focusing  when $\Omega<1$.

Our tighter constraints rule out the Planck equipartition principle
for primordial global shear and vorticity
 in its most general form \cite{Barrow}. If we consider
logarithmic decay of shear through the radiation era due to
collisionless stresses, then a rough estimate gives 
 $(\sigma/H)_{Pl}\simeq (\sigma/H)_0(1+z_{curv})(1+z_{eq})^{-1}
\{1+\ln(t_{eq}/t_{Pl})\}^{-1}$ where $z_{curv}$ ($z_{eq}$) is
the redshift of curvature (matter) domination. For the ``best'' case
of $\Omega_0=1$ we obtain $(\sigma/H)_{pl}\simeq 10^{-3}-10^{-4}$.
Our  argument applies to the most general
allowed set of globally anisotropic models: generic open or flat homogeneous, anisotropic but asymptotically
Friedmann models.
While it is possible that for more rapid decay of shear one can
allow some relic anisotropy, we conclude that any residual shear
or vorticity is constrained to be so small that it is most unlikely to have
 had a 
Planck era origin.

ACKNOWLEDGMENTS: We acknowledge conversations with John Barrow,
Janna Levin, and Martin White. We acknowledge partial support by NASA and DOE.
P.F. was supported by the
Center for Particle Astrophysics, a NSF Science and
Technology Center at U.C.~Berkeley, under Cooperative
Agreement No. AST 9120005.

\pagebreak
\pagestyle{empty}


\begin{thebibliography}{99}
\bibitem{vort}We can relate these two quantities through:
\mbox{$(\omega/H)_0=(\sigma/H)_0[2(1+h)(1+9h)]^{1/2}/(6x^2\Omega_0)$}.
\end{thebibliography}
\end{document}